\documentclass[%
 amsmath,amssymb,
 twocolumn,prl,
]{revtex4-1}

\usepackage{graphicx}
\usepackage{dcolumn}
\usepackage{bm}

\usepackage{xspace}
\usepackage{siunitx}
\usepackage{booktabs}

\newcommand{\Moller}{M\o ller\xspace}
\newcommand{\degree}{$^\circ$\xspace}

\begin{document}

\preprint{APS/123-QED}

\title{Measurement of \Moller Scattering at 2.5\,MeV}


\author{C.\,S.~Epstein}\email{cepstein@mit.edu}
\author{R.~Johnston}
\author{S.~Lee}
\author{J.\,C.~Bernauer}\altaffiliation[Present address: ]{Riken-BNL Research Center, Stony Brook University, Stony Brook, NY}
\author{R.~Corliss}
\author{K.~Dow}
\author{P.~Fisher}
\author{I.~Fri\v{s}\v{c}i\'{c}}
\author{D.~Hasell}
\author{R.\,G.~Milner}
\author{P.~Moran}
\author{S.\,G.~Steadman}
\author{Y.~Wang}
\affiliation{Laboratory for Nuclear Science, MIT, Cambridge, MA 02139}

\author{J.~Dodge}
\author{E.~Ihloff}
\author{J.~Kelsey}
\author{C.~Vidal} 
\affiliation{MIT-Bates Research \& Engineering Center, Middleton, MA 01949}

\author{C.\,M.~Cooke}
\affiliation{High Voltage Research Laboratory, Research Laboratory for Electronics, MIT, Cambridge, MA 02139}





\date{\today}

\begin{abstract}
M\o ller scattering is one of the most fundamental processes in QED, and a variety of modern experiments require its knowledge to high precision. A recent calculation considered the radiative process at low energy, where the electron mass cannot be neglected. To test the calculation, an experiment was carried out using the Van de Graaff accelerator at the MIT High Voltage Research Laboratory. Momentum spectra at three scattering angles are reported here and compared to simulation, based on our previous calculation. Good agreement between the measurements and our calculation is observed.

\end{abstract}

\pacs{Valid PACS appear here}
\maketitle



M{\o}ller (electron-electron) scattering is a fundamental, purely-pointlike process in QED, which provides an important means to test the Standard Model~\cite{E158,JLMoller}.  In addition, it is the dominant physical process in low-energy ($<$100\,MeV) electron scattering experiments. Thus, it is an important constraint in the design of electron scattering experiments that search for new physics beyond the Standard Model~\cite{DarkLight}.   Even for experiments with detectors that do not accept scattered M{\o}ller electrons, radiative M{\o}ller scattering can produce very large backgrounds.   Further, it is the basis of precision luminosity monitoring in electron scattering experiments~\cite{HERMES,OLYMPUS,Prad}.  

At low energies, the electron mass must not be neglected in calculating the M{\o}ller cross section, and we have calculated next-to-leading-order radiative corrections to unpolarized M{\o}ller and Bhabha scattering without resorting to ultra-relativistic approximations~\cite{Eps2016}.   Motivated by these considerations, we have carried out a measurement of M{\o}ller scattering at an incident electron energy of 2.5\,MeV and compared the results to a detailed simulation that uses our calculation. We note that the first experimental validation of Mott's relativistic theory of electron scattering was performed similarly at MIT by Van de Graaff, Buechner and Feshbach~\cite{Van1946}.




The experiment was carried out using the electron beam from the 3\,MV Van de Graaff electrostatic accelerator at the MIT High Voltage Research Laboratory.  The downward-going electron beam from the Van de Graaff was bent into the horizontal plane by a 90$^\circ$ bending magnet and then focused using a magnetic quadrupole doublet before being directed to the target.   The available targets were \SI{2}{\micro\meter} and \SI{5}{\micro\meter} diamond-like carbon foils from MicroMatter \cite{MicroMatter}, which were mounted on a ladder that also contained a beryllium oxide viewing screen.  The scattered electrons were precisely measured using a specially-designed focusing magnetic spectrometer with a focal plane detector designed for 1\,MeV electrons.  The electron beam current was typically in the range 30\,nA to 100\,nA and was measured using a specially-built Faraday Cup~\cite{John2018}.

The experimental apparatus was designed and fabricated at the MIT Bates Research and Engineering Center. The design consisted of a movable dipole spectrometer magnet (bending angle of 90$^\circ$ with a \SI{28}{cm} radius) and a scintillating tile focal-plane detector.  A tungsten collimator defined a square 1\degree$\times$\,1\degree acceptance.  The magnet rotated about the target along a fixed track allowing placements between 30$^\circ$ and 40$^\circ$.  

The entire beamline was held under vacuum in order to minimize multiple scattering of the low-energy electrons.  A flexible vacuum-bellows facilitated this. The electrons exited the internal vacuum chamber through a Kapton window a few centimeters from the focal plane. The main spectrometer magnet was a ``C''-magnet design, with an additional Kapton window at the back of the magnet. This allowed higher-energy elastically-scattered electrons to escape the system during the \Moller measurements, without producing too much background.



The focal plane detector consisted of a two-layer array of scintillating tiles.  The tiles were \SI{2.5}{mm} wide and \SI{0.5}{mm} thick and were made in two lengths: \SI{60}{mm} and \SI{160}{mm}.  These were made to our specifications by Eljen Technology and were diamond-milled in order to have optically-clear edges.  The material was their EJ-212, which is based on a combination of polyvinyltoluene and fluors \cite{Eljen}.  The instrumented active area was \SI{4}{cm} $\times$ \SI{15}{cm}, corresponding to 16 tiles (angle) by 60 tiles (momentum).


The light generated by the passage of the 1$-$2 MeV electrons through the scintillator was detected using silicon photomultiplier detectors (SiPMs). The SiPMs used were $2 \times 2$ mm$^2$ Hamamatsu multi-pixel photon counters (MPPC), S13360-2050VE.  These had a physical pitch of \SI{2.4}{mm}. The SiPMs were purchased in a large batch, and then sorted by breakdown voltage. Seventy-six SiPMs were chosen with extremely similar voltages, having a mean of $53.980\pm0.026$\ V (0.05\%). This allowed a single high-voltage supply to provide a suitable bias to all of the SiPMs. 

To align with the \SI{2.5}{mm} tiles, the SiPMs were rotated to an angle of 45\degree.  The tiles were read out alternately on the left and right sides, to allow the SiPMs to be spaced \SI{5}{mm} apart rather than constricting them to \SI{2.5}{mm}. They were installed on eight-channel boards that mounted directly to the side of the detector.

The amplifiers were intended to have both high gain and a fast rise-time. Each board contained eight channels to facilitate a 1:1 connection between SiPM boards and amplifier boards. They contained an on-board discriminator based on an LTC6754 comparator. An onboard eight-channel digital-to-analog converter (DAC) supplied the threshold voltages for the comparators.  Upon positively identifying a pulse, the comparator provided a digital output signal directly to the TDC. Each of the DAC's output voltages could be set individually using a serial interface. Individual timing offsets for each pair of channels were determined at the analysis stage, by histogramming the hit time separations. A \SI{5}{ns} window was chosen to define coincidences between the two detector layers, consistent with both the histograms and the intrinsic pulse rise time.



Data for \Moller scattering and electron-carbon scattering were acquired at the angles of 30$^\circ$, 35$^\circ$, and 40$^\circ$ at a beam energy of \SI{2.5}{MeV}. Individual runs of approximately sixty seconds allowed isolation of run periods with unstable beam, as necessary. The acquisition rates were sufficiently high to achieve statistical errors comparable to the systematics in less than sixty seconds.


A Geant4 simulation of the experiment was constructed, conceived to be as true to the actual physical design as possible. The magnetic field of the spectrometer used in the simulation was calculated from the SolidWorks model, using Ansys Maxwell software. The target foil, lead shielding, internal vacuum, Kapton windows, and external air gap were all included in the simulation.


The simulation bands trace out the envelope of a large number of simulations performed over the range of possible values of the beam-related parameters. Such parameters are detailed in Table~\ref{systematics}. In this way, systematic uncertainties were introduced into the comparison between data and theory.

\begin{table}[hbt]
 \begin{center}
    \caption{Selected systematics included in simulations.}
      \label{systematics}
  \begin{tabular*}{\linewidth}{l@{\extracolsep{\fill}}ll}
  	\hline\hline
    \vspace{0.5em} Measurement angle & $\pm$0.75$^\circ$ \\
    \vspace{0.5em} Beam angular spread & 1.0\degree to 1.05\degree\\
    \vspace{0.5em} Beam energy & $-75$\,keV to $+$10\,keV \\
    \vspace{0.25em} \parbox{3.5 cm}{\raggedright Effective target \\ multiple scattering} & \parbox{3.5 cm}{\raggedright Consistent with 0.25 $\times$ \SI{2.0}{g/cm^3} of 2\,\si{\micro\meter}-thick carbon, via~\cite{PDG}}\\
    \hline\hline
  \end{tabular*}
  \end{center}
  \end{table}

The measurement angle uncertainty, $\pm$0.75$^\circ$, represents uncertainty in both the beam angle and the spectrometer angle. It is, however, dominated by uncertainty in the beam angle: this is bounded primarily by the pipe diameter. 

The beam energy uncertainty is asymmetric, due to two sources of uncertainty. Occasional measurements of the beam energy determined that the accelerator's GVM readout is precise to within about 5--10\,keV. However, a recent measurement, performed by extracting the bremsstrahlung endpoint with a LaBr scintillator~\cite{Vavrek2018}, indicates that the beam energy is roughly 75\,keV below the GVM readout. With no further calibration data available, the entire range was used in the uncertainty estimation (fortunately, the effects were small). 

The beam angular spread was derived from an estimate of the beam's transverse geometric emittance. The beam spot size had been measured, in previous years, at a location corresponding to the upstream-most end of our experiment. Here, it was seen to be as small as \SI{1}{mm}, with optimal tuning. During the commissioning run, the beam spot was observed to be approximately 25 mm in diameter, \SI{2}{m} downstream of this location, at our BeO screen. From this, an angular divergence can be derived, thus providing an emittance when combined with the corresponding upstream spot size. In normal operation, the beam spot can be focused down to a diameter of approximately 3 mm at the target. Combining this with the estimated emittance yields an angular spread.  

The angular spreads due to beam emittance, and due to multiple scattering in the target, are separate yet intertwined effects.  Multiple scattering in the target is an effect that is largest at large angles, since the electrons pass through more of the target. On the other hand, effects due to the beam emittance are mostly independent of angle. By adjusting the magnitude of these two effects, the simulations were matched to the data. This involved omitting simulated spectra that were inconsistent with the data. The data indicate a level of multiple scattering consistent with one-fourth what would be expected for a \SI{2}{\micro\meter} target at a density of \SI{2.0}{g/cm^3}. This could indicate that the target is either thinner than expected, less dense, or that Geant4 is not handing multiple scattering accurately for such a thin material. Unfortunately, the foil's mounting hardware and it's location in the vacuum system prohibit a direct verification of its thickness. An additional beam angular spread of approximately 1.0\degree to 1.05\degree is consistent with both the data and the emittance estimation. Significantly different values of the target thickness and the beam angular spread were inconsistent with the acquired spectra. 


A necessary component of the analysis is the conversion between momentum and hit position on the detector. Being located on the focal plane, this conversion should be approximately, although not exactly, linear. This calibration was performed using the elastically scattered electrons, which have a uniform momentum. Rather than stepping the electron energy and extracting the calibration, the magnetic field of the spectrometer magnet was stepped, as this has an equivalent effect. 

As a result, the elastic peak was swept across the focal plane detector. The position was determined by a Gaussian fit to the top of the peak. The extracted position of the elastic peak (tile number) was determined as a function of magnet current. 
The magnet current was then translated to an effective electron momentum, and the data fit to a third-order polynomial. An orthogonal distance regression was used, owing to the presence of both tile coordinate $x$ and $y$ uncertainties. 

The same procedure was repeated with the simulated detector, in order to extract a mapping for use on the simulated data. This was done in order to help mitigate effects resulting from differences between the simulated and real magnetic field. By performing the calibration twice, and using the simulated map for simulated data, and the experimental map for real data, these discrepancies can be largely cancelled. 


The effects of light attenuation were clearly visible in the data. This attenuation was modeled, fit, and then the data were corrected. A double-exponential model was used as a starting point for the model. This contains two terms: one for light attenuated as a result of internal reflection, and a term for light attenuated in the bulk material. The bulk attenuation length is quoted by the manufacturer as \SI{2.5}{m}, which indicates practically constant (and negligible) attenuation on the relevant short length scales. The rate $R$, of hits along a strip as a function of distance $x$, was thus parameterized as $R(x) = \exp(-x/l) + C$, with free parameters $l$ (reflective attenuation length) and $C$ (bulk offset). The overall scale was fixed. 

To extract the values of these parameters, two splines were fit around each edge of the detector: one on even tiles, and one on odd tiles (corresponding to opposite-side readouts). With a proper correction for light attenuation, these splines should converge. The parameters of the correction model were then fit in order to minimize the difference between the splines. 

The detector efficiencies were extracted from the data using an iterative unfolding method. The $X$ and $Y$ tiles were treated separately in each iteration. To calculate each cycle's $X$-tile efficiencies, splines were fit to the rates of the $Y$ tiles. Then the tile efficiencies were fit in order to minimize the sum, at every point, of the squared deviations of the splines from the data. The same method was used to find the $Y$-tile efficiencies, by fitting splines along the $X$ tiles. The end result efficiencies were determined by multiplying the intermediate efficiencies of all of the iterations. The algorithm converged relatively quickly, in approximately 25 or fewer iterations. 

The iterative unfolding method is useful when the underlying ``true'' data can be well-represented by splines. To that end, spline-induced bias is minimized when the data are as flat and smooth as possible. Such ``flat'' spectra were generated by scanning the magnet current to methodically move the M{\o}ller peak across the detector. Efficiencies were reconstructed from this relatively flat data set, and then applied to the real data of interest. Some bias is unavoidable based on the validity of the assumptions, but it is ideally small in the most-important central regions of the detector. It is also important to note that this method can only provide the relative efficiency between the tiles, not the absolute efficiency. Likewise, it cannot account for long-range structure in the detector efficiency, only short-range tile-by-tile variation. 

The efficacy of this reconstruction method was evaluated by using a toy model. Fake efficiencies were applied to a flat data set, which was then fed through the reconstruction algorithm. The efficiencies were drawn from a normal distribution with a mean of 1 and width of 0.05: the efficiency parameters were reconstructed to roughly $\pm$10\%. As a result, the error bars on the presented data points consist of both uncertainty from statistics, and that resulting from an estimated $\pm$10\% uncertainty in the efficiency parameters. 

\begin{figure}[htb]
  \centering
  \includegraphics[width=\linewidth]{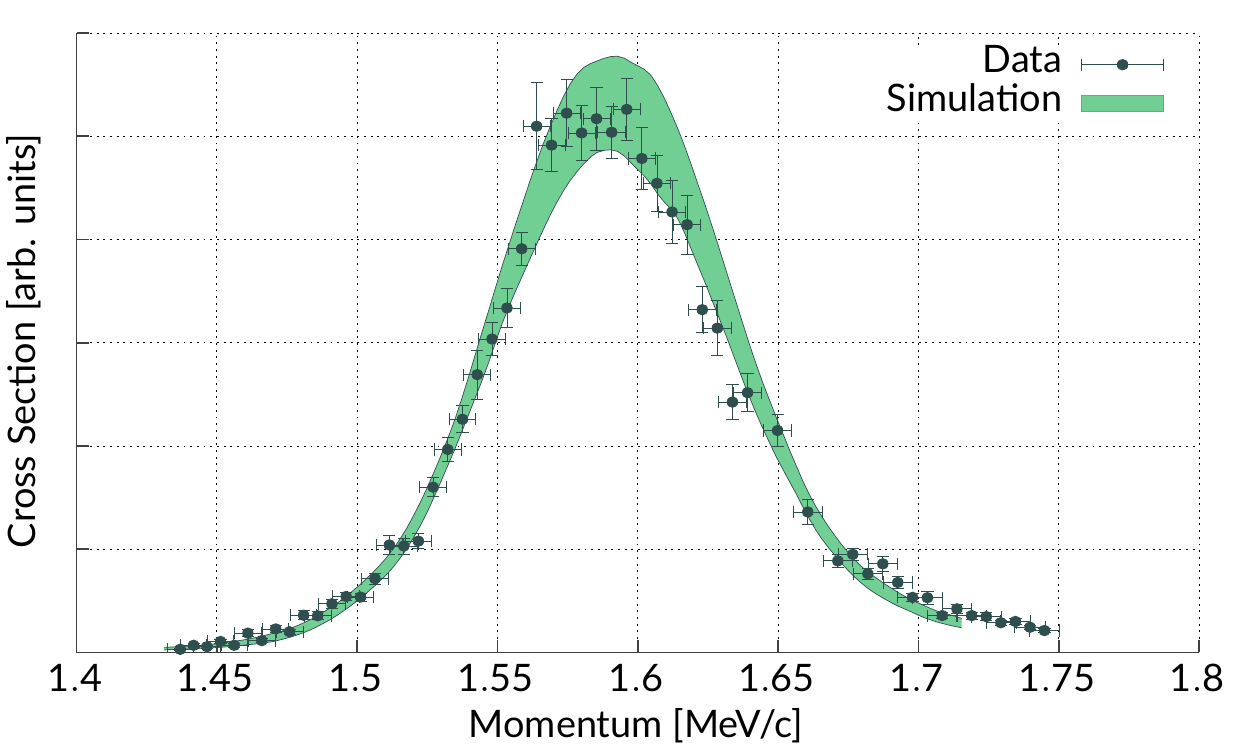}
  \caption{Yield of scattered electrons vs.\ momentum compared to simulation at an electron scattering angle of 30$^\circ$.}
  \label{comp30}
\end{figure}
\begin{figure}[htb]
  \centering
  \includegraphics[width=\linewidth]{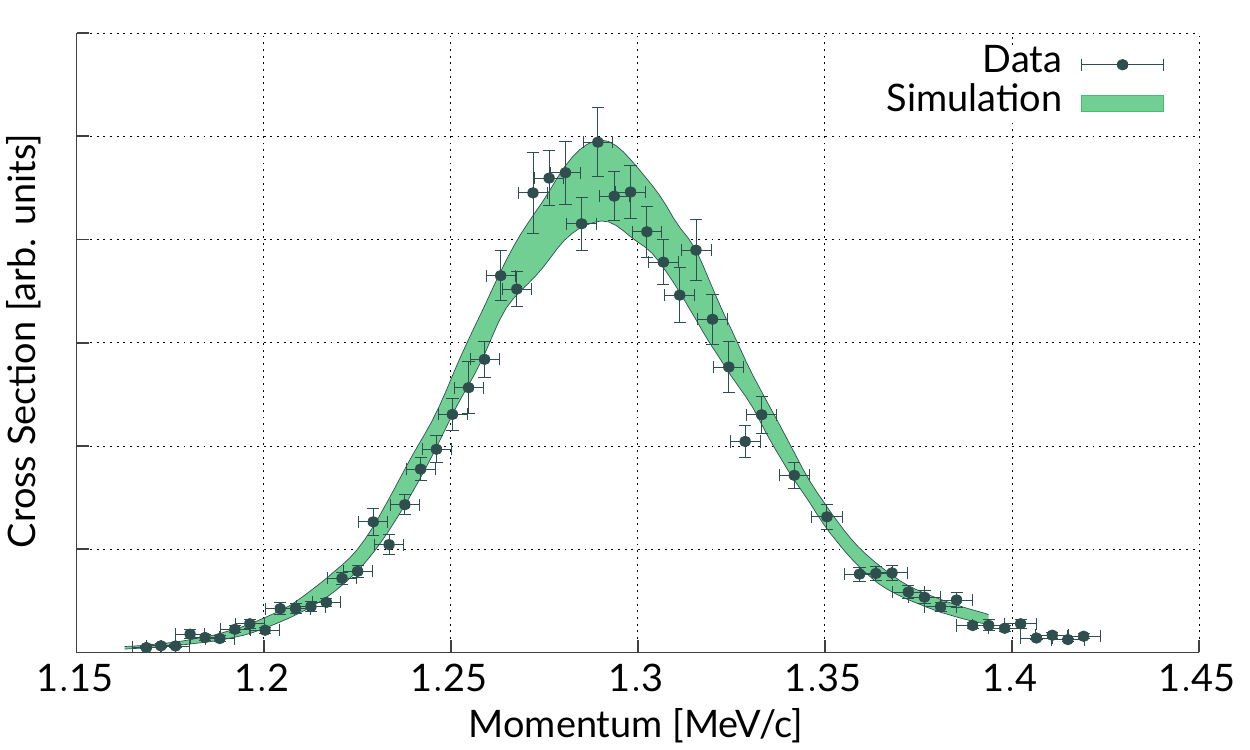}
  \caption{Yield of scattered electrons vs.\ momentum compared to simulation at an electron scattering angle of 35$^\circ$.}
  \label{comp35}
\end{figure}
\begin{figure}[htb]
  \centering
  \includegraphics[width=\linewidth]{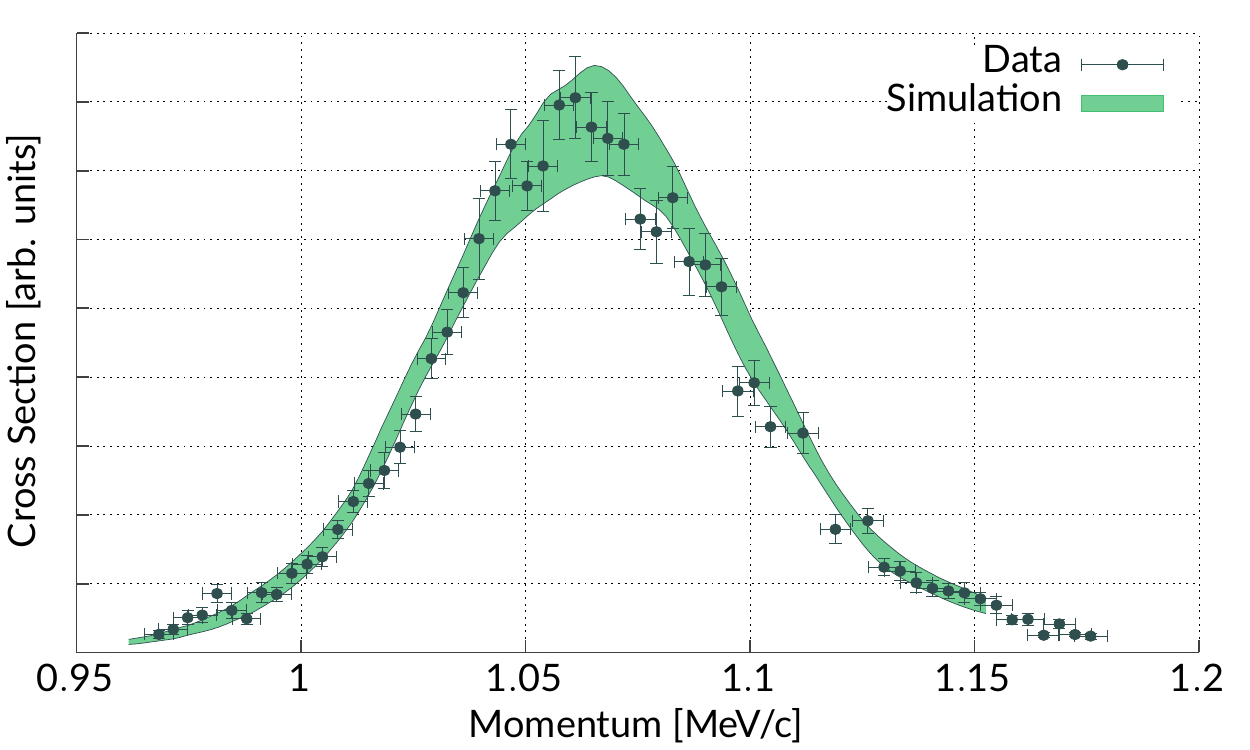}
  \caption{Yield of scattered electrons vs.\ momentum compared to simulation at an electron scattering angle of 40$^\circ$.}
  \label{comp40}
\end{figure}
\begin{figure}
  \centering
  \includegraphics[width=\linewidth]{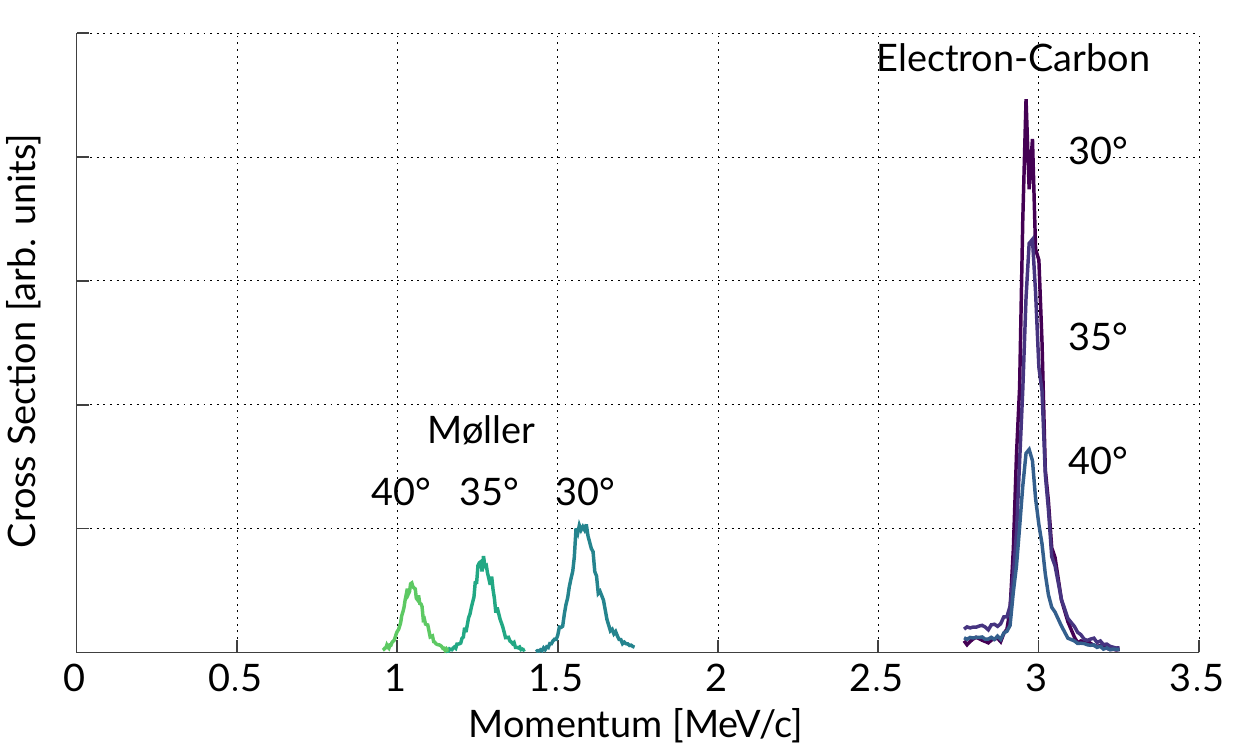}
  \caption{Composite spectrum from several different measurements of the yield of scattered electrons vs. momentum at an electron scattering angle of 30$^\circ$, showing the M{\o}ller (electron-electron) and elastic (electron-carbon) peaks.}
  \label{fullscale}
\end{figure}

Figures~\ref{comp30}, \ref{comp35}, and \ref{comp40} show a comparison between the extracted M{\o}ller spectra and that reconstructed from a complete radiative simulation. The data have been scaled vertically to match. Small (sub percent-level) horizontal offsets were added to the data in order to optimize the overlap. These small offsets are consistent with uncertainties resulting from magnet hysteresis and the intrinsic accuracy of the power supply. The composite momentum spectrum is shown in Fig.~\ref{fullscale}, demonstrating the relative positions of the electron-carbon and M{\o}ller peaks. 


The M{\o}ller data show good agreement with the predictions of the new radiative theoretical calculation. The shape of the spectra at 30$^\circ$, 35$^\circ$, and 40$^\circ$ are well-described: this was the primary goal of this experimental effort. Some deviations from the general trend are seen in the 30$^\circ$ comparison, although these fall within the uncertainties. Future iterations of this experiment should be able to improve upon the precision of this measurement. 

In summary, we have carried out a measurement of M{\o}ller scattering at an electron energy of 2.5\,MeV on a carbon target.
We have developed a focusing spectrometer and focal-plane detector using modern scintillator tiles and readout optimized for detection of 1\,MeV electrons.  The measured M{\o}ller spectra at 30$^\circ$, 35$^\circ$ and 40$^\circ$ are in good agreement with a simulation that is based on a calculation that includes the finite value of the electron mass, as well as all low-energy interaction processes of the electrons as they pass through material.  This work validates the current understanding of M{\o}ller scattering at energies where the electron mass cannot be neglected, which is a significant constraint in the design of high-intensity, low-energy electron scattering experiments.   

This research was supported by the National Science Foundation under MRI Award No. 1437402, as well as the DOE Office of Nuclear Physics under grant No. DE-FG02-94ER40818, and a DOE National Nuclear Security Administration Stewardship Science Graduate Fellowship program under grant number DE-NA002135.


\end{document}